\documentclass[prd,amsmath,amssymb,nofootinbib]{revtex4} 
\usepackage{graphicx} 
\usepackage{amsmath}
\usepackage{amsfonts,amsbsy}
\usepackage{amssymb}
\usepackage{breqn}

\newcommand{\be}{\begin{equation}}
\newcommand{\ee}{\end{equation}}
\newcommand{\bea}{\begin{eqnarray}}
\newcommand{\eea}{\end{eqnarray}}

\def\lsim{\mathrel{\rlap{\lower4pt\hbox{\hskip1pt$\sim$}}
    \raise1pt\hbox{$<$}}}                % less than or approx. symbol
\def\gsim{\mathrel{\rlap{\lower4pt\hbox{\hskip1pt$\sim$}}
    \raise1pt\hbox{$>$}}}                % greater than or approx. symbol

\begin{document}
\title{Renormalization group ontology in quantum foundations: a non-interacting spin-0 toy model}

\author{Gary Kapilevich}
\email{gakapilevich@loyola.edu}
\affiliation{Department of Physics$,$ Loyola University Maryland$,$\\
 4501 North Charles Street$,$ Baltimore$,$ Maryland$,$ 21210$,$ USA}

\begin{abstract}
A realist, anti-psi interpretation of quantum field theory is introduced, based on an expanded ontology that was inspired by the renormalization group. A goal will be to work with a classical non-interacting scalar field ``toy model” in the canonical ensemble and show that an observer in this expanded ontology, under certain limiting conditions, will see a QFT. The inverse renormalization group (IRG) will be key in both interpreting this expanded ontology and furnishing novel examples of QFTs. This will lead to a definition of Planck’s constant that flows as we head towards a fixed point, is expressed in terms of a temperature, and depends on the observer being studied in this expanded ontology. Our hope is that this approach will lead both to insights into quantum foundations and to prescriptions for developing new QFTs.
\end{abstract}

\maketitle

\section{Introduction}
\label{section: introduction}

We put forward an interpretation of QFT based on the renormalization group. Let’s start by briefly comparing this interpretation to a few others. We take some inspiration from Spekkens’ toy model\cite{PhysRevA.75.032110}: not as an interpretation of QM, but as a model that replicates many quantum mechanical phenomena by starting from states with incomplete information. A line from the introduction is particularly relevant:
\begin{quote}
This suggests that one would obtain a better analogy with quantum theory if states of complete knowledge were somehow impossible to achieve, that is, if somehow maximal knowledge was always incomplete knowledge.
\end{quote}
We also draw inspiration from Rovelli’s relational QM\cite{rovelli1996}: that states should always be defined relative to some observer. These two notions – information should be withheld in QM, and states should be defined relative to observers with different information – are critical to the present attempt. 

But we think more is needed. In many worlds, a multiverse of worlds (depending on the version\cite{quantum6020011}) is built off an expanded ontology of the wavefunction. We also want to build an expanded ontology of the universe: to enlarge models of it beyond bounds of the Planck length on the one hand and the observable universe on the other. But RG ontology is, by design, realist and \emph{anti-psi}. This is potentially time saving, because unlike, say, Bohmian and dynamical collapse interpretations, we don’t have to supplement the unitary evolution of the wavefunction with some other dynamics (see sources for a review\cite{Struyve_2010}\cite{RevModPhys.85.471}). But if we do not have recourse to $\psi$, what do we draw on for this enlarged worldview?

We are trying to motivate a model where one observer’s UV field (where the UV modes have been integrated out with the renormalization group) is a different observer’s IR field (where the IR modes have been integrated out). We dream of a model where someone’s high energy physics is someone else’s observable universe. That observers can \emph{live}, completely, in the high energy modes another observer has integrated out, and vice versa. The renormalization group, or, for observers who integrate out low energy modes, the \emph{inverse renormalization group}, is key. We set up an \emph{RG hierarchy} of modes, and observers integrate out different modes in the hierarchy based on the information they have of it. With our demand of a self-similar fixed point in section \ref{subsec: QFT_postulates}, we are using the renormalization group to set up a sort of ``matryoshka universe”.

Importantly, we will argue that there is a payoff to modeling the universe in this way: emergent QFT. This is based off an idea about fields related to each other through analytical continuation of the time: we have in mind that a QFT in 4D Minkowski spacetime is related to a Euclidean classical field in 4D \emph{space} (through Wick rotation and surreptitious replacement of $\hbar$ with a temperature). We think both should be viewed as limits of a non-relativistic, three-dimensional, classical field. This field will live somewhere in the RG hierarchy, with high \emph{and} low energy cutoffs, and will see a QFT when the cutoffs obey \emph{radiation zone limits} and when the field is at a self-similar fixed point. Planck’s constant will be defined proportional to the temperature at the fixed point.

We work with the non-interacting scalar field in this paper, which is very constraining! For example, it is important to our interpretation that $\hbar$ and $c$ are renormalized, but they obviously aren’t in the non-interacting theory! Instead, we make speculative arguments about their renormalization in section \ref{sec: toy model}, consistent with the present model but for purposes of future generalization. We try to lay out the ideas of how all this might work in this paper in a simplified context and hope for their future application to fields with interaction and spin.

The structure of this paper is as follows. We discuss philosophy and postulates in section \ref{sec: philosophy}. RG ontology is introduced in section \ref{subsec: RG_intro}; the postulates are introduced in section \ref{subsec: QFT_postulates}; and we take on additional topics like Wigner’s friend and the measurement problem in section \ref{subsec: foundations}. The actual non-interacting scalar field model, our primary application, is the job of section \ref{sec: toy model}. We start with the 3D non-relativistic classical field in subsection \ref{sec: classical physics}; renormalize in subsection \ref{subsection: postulates 1 and 2}; discuss prospects for the flow of $\hbar$ and $c$, along with the Gaussian fixed point, in subsection \ref{subsection: hbar, c, and postulate 3}; discuss the propagator in subsection \ref{subsection: propagator}; and commutation relations in subsection \ref{subsection: commutation relations}. Finally, in section \ref{sec: RG}, we leave familiar terrain (where we know how the QFT should look) and start talking about observers who see an IR field. We first discuss Wigner’s friend and the classical limit in section \ref{subsec: classical limit}; take on the IR field in section \ref{subsection: IR QFT}; discuss what we’ll define in the next section as the active and passive pictures, in section \ref{subsection: Active/Passive}; and use RG and IRG for a single observer integrating out very many high and low energy modes in section \ref{subsection: Alice's renormalization group}. We end that section by speculating about use of the renormalization group as a group, and not a semigroup.

%---------------------------------------------------------------------------------------------------------------------------------------

\section{Philosophy of the Renormalization Group Ontology}
\label{sec: philosophy}

%---------------------------------------------------------------------------------------------------------------------------------------

\subsection{The RG Hierarchy}
\label{subsec: RG_intro}

To set the scene, imagine that we are renormalizing a classical field, for which we define a sequence of modes, 
\be
\Lambda’\ll \Lambda_1\ll \Lambda_2 \ll \Lambda_3 \ll \Lambda_{n-1}\ll\Lambda_n\ll\Lambda~,
\label{eq: hierarchy}
\ee
where the limits of eq.~(\ref{eq: hierarchy}) will be explained in section \ref{subsec: QFT_postulates}. 
We situate an observer, Alice, at $\Lambda_A\approx\Lambda_1$. Say that Alice wants to 
describe a field at $\Lambda_2$ but does not have access to any higher energy modes. She 
decides to apply RG with the following familiar steps: she starts with a Lagrangian with 
very many interaction terms; she defines her bare Lagrange coefficients at the 
high energy/UV cutoff $\Lambda$; she integrates out modes in the interval $(\Lambda_2 , \Lambda)$; 
she then computes renormalized Lagrange coefficients, many of which will become irrelevant at a fixed point if the 
field she’s using is renormalizable. This is the standard way that things are done in effective field theory 
going back at least to Wilson’s seminal paper\cite{WILSON197475}. We’ll call the result a 
small scale/UV field, after the UV modes that have been integrated out.

We now place a second observer, Bob, at $\Lambda_3$. From Alice’s perspective, 
Bob sits at one of the UV modes Alice has integrated out in her attempt to describe a field 
at $\Lambda_2$. But perhaps Bob doesn’t have access to low energy modes at 
$\Lambda_1$ (where Alice is), views {\em them} as very complicated, and decides to integrate 
them out in describing a field at $\Lambda_2$: he defines his bare Lagrange parameters at $\Lambda’$, 
applies the {\em inverse renormalization group (IRG)} to integrate out modes in the 
interval $(\Lambda’,\Lambda_2)$, and calculates renormalized parameters. We’ll call the result a 
large scale/IR field. Both Alice and Bob are describing the same object; they are situated differently 
in the {\em hierarchy} of eq.~(\ref{eq: hierarchy}), have access to different pieces of 
information, and so renormalize accordingly.

What does it mean to situate an observer like Bob in modes that another observer, Alice, has integrated 
out in her application of RG? We imagine that Bob is doing a type of ``quantum cosmology”. To visualize this, 
we’ll use a thought experiment which we’ll call the {\em IR observable universe}. We imagine Bob’s observable 
universe as surrounded by some complicated large scale (low energy) physics that an observer on 
Bob's planet does not have access to. Bob then applies IRG to get a 
large-scale theory of (what is to them) the observable universe. Comparing Bob and Alice, the thought 
experiment implies that Bob’s ``IR observable universe” is Alice’s ``small scale UV field”. Please note: we are {\em not} 
suggesting that quantum cosmology should be done with a non-interacting scalar field (there is no general relativity in this paper)! 

More generally, we imagine placing observers up and down the hierarchy of eq.~(\ref{eq: hierarchy}) and 
allowing them to study fields at an arbitrary mode by using either RG or IRG. 
This has the effect of ``decoupling modes from complexity": we say that the 
original field, defined on $(\Lambda',\Lambda)$ with many interacting terms, 
is ``complicated", but that any given perspective that has integrated out modes, even perspectives that have kept the UV ones, need 
not be complicated. Let’s emphasize this intuition with a statement about RG ontology:\\ \\
{\em Any mode of eq.~(\ref{eq: hierarchy}) can have an observer situated there, 
and any resulting perspective might be either complicated (with the full regalia of Lagrange parameters) 
or simple (with most of the parameters becoming irrelevant) depending on 
observer’s place within the hierarchy and the modes they do not have access to and have integrated out.}\\

The inverse renormalization group has some recent applications 
as disparate as fluid dynamics and string theory\cite{Nastase_2020,GAWEDZKI1997123}. We will use it in 
the context of a toy model of the non-interacting scalar field, in which modes don't mix in the 
Hamiltonian, so that $H[\phi]=H[\phi^{-}]+H[\phi^{+}]$, where $\phi^{-}$ is over the $(\Lambda',\Lambda_1)$ modes while $\phi^{+}$ is over $(\Lambda_1,\Lambda)$. In sections \ref{sec: toy model} and \ref{sec: RG}, we will use RG to integrate 
our the high energy modes in, say, the interval $(\Lambda_1,\Lambda)$, and will sometimes use IRG to integrate out the 
low energy modes in the interval $(\Lambda',\Lambda_1)$. We will sometimes find it useful to write a single expression that includes both possibilities:
\be
Z_{\mp}=\int\mathcal{D}\phi^\mp e^{-\beta H[\phi^\mp]}\int\mathcal{D}\phi^\pm e^{-\beta H[\phi^\pm]}~,
\label{eq:mode seperation}
\ee
where the minus on the partition function describes a small scale field for which the UV modes have been integrated out 
while the plus on the partition function describes a large scale field. Eq.~(\ref{eq:mode seperation}) might describe the physics of a single field from the perspective of two
different observers (one who integrates out the UV modes and the other the IR modes). We'll call this the {\em passive picture}.
One can also interpret eq.~(\ref{eq:mode seperation}) as describing two distinct fields that a single observer models by starting
with the general theory on $(\Lambda',\Lambda)$ in each case and integrating either UV or IR modes. We'll call this the {\em active picture}.

At the end of subsection \ref{subsection: Alice's renormalization group}, we will discuss the possibility of using the renormalization group as a {\em group}, not a semi-group. We will imagine an observer
like Alice that doesn't have access to some subset of a field's high {\em and} low energy modes. 
We speculate about a model where the inverse of using RG to integrate out high energy modes is to then use IRG to integrate
out low energy modes. We emphasize that in thinking about using RG as a group in this way, we 
are {\em not} trying to put back modes that have already been integrated out (as is being attempted, for example, in
some recent work with computational algorithms\cite{PhysRevLett.128.081603}).

Should we call this RG ontology “infinitely divisible”? This would mean starting with 
eq.~(\ref{eq: hierarchy}) but then taking the limits $n\rightarrow\infty$, $\Lambda\rightarrow\infty$, and $\Lambda'\rightarrow 0$. In QFT, the high energy cutoff can be removed, for example, in asymptotically free theories \cite{Srednicki_2007}. A reason usually given for not having an infinite cutoff is hubris: we should not imagine that QFTs can describe physics at arbitrarily high energy modes. The logic of this paper is slightly different, because we see no reason why we should not put an observer into one of these high energy modes and imagine that they might even see renormalizable physics there if they integrate out very many low energy modes. So we agree that infinite cutoff theories are unphysical but we want to study ``very large” theories in classical statistical mechanics, large enough that nested observers can be placed all along an RG hierarchy and the limits of eq.~(\ref{eq: hierarchy}) apply. Of course, imagining an ontology of nested observers is its own kind of hubris. Our hope is that the interpretation of QFT we will furnish 
out of the RG ontology of this section will be useful enough to justify any perceived excess. 
In the rest of this paper, we will describe this QFT interpretation.

%---------------------------------------------------------------------------------------------------------------------------------------
\subsection{The Three Postulates of (Non-Interacting Scalar) QFT}
\label{subsec: QFT_postulates}

It’s an old observation that if you start with a QFT in four-dimensional Minkowski \emph{spacetime} and do a Wick rotation, the resulting Euclidean QFT corresponds to a classical field theory in four dimensional \emph{space}, but with $\hbar$ instead of a temperature:
\be
Z_{QFT}=\int\mathcal{D}\phi~e^{\frac{i}{\hbar}S_{QFT}}\rightleftharpoons \int\mathcal{D}\phi~e^{-\beta\int\mathrm{d}^4 x\mathcal{H}}~.
\label{eq:EuclideanPartition}
\ee
Say that someone came along and claimed that the \emph{physics} was with the classical field in four-dimensional space. They’d face a slew of questions\cite{TongStatField}. What \emph{is} a four-dimensional spatial classical field theory? How can you get a quantum field theory in a vacuum, with an $\hbar$, from a classical field theory with a temperature in the canonical ensemble? What if the classical Hamiltonian were complex after a Wick rotation? And if you go from a classical field to many classical particles, wouldn’t these particles behave deterministically and not quantum mechanically?

The last question suggests a better approach: why not look for a mechanism that makes a particle in a classical ensemble impossible to isolate? Let’s try to do this with the renormalization group. Start with a \emph{three dimensional, non-relativistic} classical field in the canonical ensemble. Give the field cutoffs - $\Lambda’$ for the low energy cutoff and $\Lambda$ for the high energy cutoff – and integrate out modes up through $\Lambda_1$. Then apply the following three postulates.
\begin{enumerate}
\item Integrate out the high energy modes to $\Lambda_1$.
\item Take the {\em radiation zone} limit: $\Lambda' \ll \Lambda_1 \ll \Lambda$.
\item After the limits of postulate 2 have been taken, allow RG flow towards a {\em self similar fixed point}.
\end{enumerate}
We’ll call these P1, P2, and P3. Claim: we’ve started with a classical field and ended up with a \textbf{small scale QFT}. The claim is also about what it means to have two fields – a Euclidean classical field in 4D space and a Minkowski QFT in 4D spacetime – related to each other through analytical continuation of the time. They are both limits of a non-relativistic 3D classical field theory, the limits placing the classical field somewhere in an RG hierarchy.

The procedure described should furnish an expression for Planck’s constant in terms of the temperature at the fixed point, which shouldn’t be surprising: a recent work on ``effective quantum mechanics” for biological systems, for example, has several expressions for ``effective Planck’s constants”\cite{HUBSCH2024169641}. We’d also like both Planck’s constant and the speed of light to be renormalized as we head towards the fixed point with P3, which should dress each with the integration bounds. For most of this paper, we will hem to the following \emph{Planck’s constant/speed of light renormalization heuristic}:\\ \\
{\em Planck’s constant, $\hbar$, should be proportional to the temperature at the fixed point. We define a bare speed of light, $c_0$, and a renormalized speed of light, $c$.  Both $\hbar$ and $c$ should decrease as modes are integrated out. This should be the case whether RG or IRG is used. }\\ \\
We now discuss these requirements with the emergence of quantum mechanics and locality in mind.

%---------------------------------------------------------------------------------------------------------------------------------------

\subsubsection{A Discussion of the Postulates.}
\label{subsec: discussion of the postulates}

Had we integrated out the low energy modes instead of the high energy ones of postulate 1, we might still have gotten a QFT:
the idea is that {\em anytime} you integrate out ``effectively infinite” modes (with ``infinite” defined by the limits of postulate 2), you might get a QFT. In the IR observable universe
thought experiment, Bob doesn’t have access to his observable universe’s large-scale modes, integrates them out, and sees a QFT if postulates 2 and 3 hold. 
We emphasize that this would be a QFT ``from above”, created because of Bob’s ignorance of the low energy modes and {\em not} because of any small scale QFT Bob separately observes. 
We include postulate 1 to lay out the procedure for small scale QFT and to emphasize that there are other quantum mechanical theories, the exploration of which will be the job of section \ref{sec: RG}.

We name postulate 2 after the “radiation zone limit” of a radiating dipole in classical electrodynamics, though we do not take the metaphor farther than the name. Postulate 2 and 3 are meant to fulfill the requirement for quantum mechanics we mentioned at the beginning of this section: we study a field on the interval, before rescaling, $(\Lambda’,\Lambda_1)$, and we try to take into account influences on this field from the high energy regime beyond $\Lambda_1$. Because of the limits of P2, these influences are ``very far away”, so an observer in the RG hierarchy situated around $\Lambda_1$ will ``never” be able to access these far away modes. Because we are at a self-similar fixed point, curtesy of P3, these impossible to access modes still influence the field.

Placing ourselves somewhere in the RG hierarchy gives us a low energy cutoff. So rescaling the UV theory will no longer reproduce the original bounds. 
For example, the low energy cutoff for the UV theory in section \ref{sec: toy model} will change 
from $\Lambda'$ to $\Lambda'_{-}$, where $\Lambda'_{-}=\Lambda' \frac{\Lambda}{\Lambda_1}$. The 
renormalized low energy cutoff now “wanders”, so that postulate 2 must be applied before postulate 3 or 
else we get a trivial theory. To get a renormalized theory we can compare to the original, we use the following {\em cutoff rescaling criteria}:\\ \\
{\em Say that the cutoffs of a field are $(\Lambda',\Lambda)$ before renormalization and $(\Lambda'_{\mp},\Lambda_{\mp})$ after. 
We require that one of the new cutoffs be the same as one of the original cutoffs. We also require 
that the new bounds satisfy $\Lambda'_{\mp}\ll\Lambda_{\mp}$, just like the original cutoffs after application of postulate 2.}\\

We’ll spend most time on the ``Gaussian fixed point”, as this gives us the correct behavior for the propagator with a divergent correlation length for the UV field. Gaussian fixed point is in quotes because we now have a low energy cutoff (not 0), so the idea will be to fine tune the bare mass to hit the rescaled low energy cutoff, $\Lambda’_{-}$. We’ll also look at the fixed point at infinity and a ``massless fixed point” for the IR field, where the latter only means that, after rescaling, $\mu_{+}\ll\Lambda’_{+}$. Not all of these fields will have a divergent correlation length as required by P3. 

Finally, let’s give a schematic similar to eq.~(\ref{eq:EuclideanPartition}) for the Hamiltonian, which reflects our conviction that limits on a 3D nonrelativistic classical field can produce a 4D Minkowski QFT:
\be
\lim_{\mathrm{P2},\mathrm{P3}}\left(-H[\phi^{-}]\beta\right)=\frac{i}{\hbar}S_{\mathrm{QFT}}[\phi^{-}]~.
\label{eq: classical/quantum equivalence}
\ee
You first take the radiation zone limit of the UV theory on the left side. You then allow RG flow with the limit in P3, so that $\xi_{-}=\frac{\Lambda}{\Lambda_1}$ is allowed to increase. On the right side we have a ``QFT action", $S_{\mathrm{QFT}}$, in Minkowski space with a Planck’s constant appropriate for the field being studied.

%---------------------------------------------------------------------------------------------------------------------------------------

\subsubsection{Emergent QFT.}
\label{subsec: emergent qft}

In this paper, we look to reproduce a quantum field theory in the vacuum, but we start in the canonical ensemble with a temperature. The idea is that we need to be at a fixed point at P3 with a large correlation length, so set the temperature accordingly. If we were working with $\phi^4$, we might expect that in a mean field approximation this would be the critical temperature. But this paper is about the non-interacting field, so we will instead argue (in section \ref{subsection: hbar, c, and postulate 3}) that the temperature should be proportional to the small constant $\epsilon$. That argument will hinge on the renormalization (or lack thereof, for the non-interacting field) of Planck’s constant, into which the fixed-point temperature will be absorbed.

We want $\hbar$ to decrease with RG or IRG because we want decoherence to occur if we integrate out additional modes beyond the fixed point. The idea is that as more modes are integrated out, the field over the remaining modes has an \emph{effective Planck’s constant}, $\hbar_{\mathrm{eff}}$, smaller than $\hbar$, allowing us to use the saddle point approximation in the classical limit. So the more modes we integrate out, the less information we have on the field but the easier it is to ignore fluctuations around the saddle point. 

In section \ref{subsection: hbar, c, and postulate 3}, we will propose a mechanism for renormalizing $\hbar$ based on wavefunction renormalization. There, we discuss how to rescale the field so that we get expressions for $\hbar$ and $c$ proportional to the renormalized parameter $\gamma$. There is nothing systemic about this here, and an ultimate schema for the renormalization of $\hbar$ and $c$ will have to wait for the interactive scalar RG ontology theory.

Whether the classical limit is approached or not, we call all observers in the RG hierarchy who have integrated out very many modes with P2 \emph{non-deterministic observers}. We say that a non-deterministic observer doesn’t have information about the modes they have integrated out. \emph{Deterministic observers}, on the other hand, haven’t used RG to integrate out any modes at all. The deterministic observer see classical statistical mechanics in the canonical ensemble while the non-deterministic observers see an emergent QFT.

How has locality emerged from non-relativistic classical physics in the canonical ensemble? The deterministic observer sees a Euclidean spacetime with the speed $c_0$ associated with the overall field. The non-deterministic observer, having integrated out very many modes up through $\Lambda_1$, only has access to the renormalized speed of light, which, by our heuristic, has decreased to $c<c_0$. We assume that any additional modes the non-deterministic observer integrates out (in the pursuit of, say, the classical limit) will not significantly renormalize $c$ that much more, because the distance between these additional modes and $\Lambda_1$ will pale in comparison to the distance between $\Lambda_1$ and either cutoff. So the non-deterministic observer sees physics at $c$ over the vast swath of modes they \emph{do} have access to. 

It’s as if, as the non-deterministic observer coarse grains, they stop ``seeing” the smaller scales that the deterministic observer still sees. Perhaps things at those smaller scales move faster than things the non-deterministic observer sees. But such scales have disappeared for the non-deterministic observer, their effects only felt through a Minkowski spacetime that the non-deterministic observer is forced to inhabit.

%---------------------------------------------------------------------------------------------------------------------------------------

\subsection{Select Topics in Quantum Foundations}
\label{subsec: foundations}

%---------------------------------------------------------------------------------------------------------------------------------------

\subsubsection{Anti-Psi-Ontology}
\label{subsec: antipsi}

Consider again the IR observable universe thought experiment, where 
Bob studies his observable universe by integrating out its low energy modes in accordance with P2 and P3 of section \ref{subsec: QFT_postulates}. Bob's observable universe would have a wavefunction that 
evolves unitarily; Bob might decide to measure some property of the observable universe at time $t$ after a big bang; 
the wavefunction of the observable universe would then collapse upon measurement; Bob 
might then try to measure the same property at $t'$ and might get the same or different result, depending on if
he has taken a generalized measurement, etc... 

Bob's observable universe can interact with the low energy modes 
very far from both it and Bob, and Bob himself can also interact with his observable universe. But it seems clear that, to a deterministic observer,
the state of Bob's observable universe at time $t$ is due to an 
interaction between it and the low energy modes and not due to any 
interaction between Bob and his observable universe. The latter only reveals information 
to Bob, who doesn’t have access to the low energy modes.

We then invoke “self-similarity” to argue that the same holds for Alice studying 
a UV field. She only has access to a field’s low energy modes, integrates out the high 
energy ones, and, provided the postulates hold, sees a quantum mechanical theory 
with a collapsing wavefunction. But just like Bob, we assume that a collapsing wavefunction 
only reveals information to Alice, and that the state of the field she’s observing is determined by an 
interaction with high energy modes that are far from Alice and that she doesn’t have access to.\\

%---------------------------------------------------------------------------------------------------------------------------------------

\subsubsection{The Measurement Problem}
\label{subsection: the measurement problem}

One motivation for the self-similarity requirement in P3 is the collapse of the wavefunction for an entangled pair of electrons, which occurs regardless of how far the pair are and so suggests a divergent correlation length for the UV field. RG ontology is an anti-psi interpretation, so we don’t view the collapse as determining the eigenstate of the entangled pair upon measurement. Instead, collapse reveals information to observers, and the observed eigenstate  is determined by interactions with the integrated-out UV or IR modes. 

We illustrate what happens when observers in an RG hierarchy take measurements by giving Alice an entangled electron. We give the other electron to Carl, who observes UV fields like Alice, and send Alice and Carl to opposite sides of their galaxy. We imagine Alice taking a measurement of her electron: her wavefunction of the EPR pair collapses and she sees the EPR pair in one of its eigenstates, say the one in which Alice’s electron is spin up. Carl takes a measurement of his electron: his wavefunction for the EPR pair collapses and he sees his electron as spin down. We would also like to introduce a third observer to the story, Bob, who sees the EPR pair as some sort of IR field: a large-scale field of the order of Bob's observable universe. Like Alice and Carl, we imagine that Bob takes a measurement of one of the EPR pairs, at which point Bob’s wavefunction collapses and he sees a definite spin for the electron that he is studying. 

Alice, Bob, and Carl are non-deterministic observers; to a fourth, deterministic observer, the entangled pair of electrons make up some sort of classical field, and such an observer would see everything behave deterministically, subject to whatever locality this observer imagines the classical field obeys. Alice, Bob, and Carl each have limited information on the entangled pair. From Alice and Carl’s perspective, the state of the entangled pair is completely determined by an interaction between the entangled pair and UV modes which are ``very far” from Alice and Carl courtesy of postulate 2. When either makes a measurement, their wavefunction instantaneously collapses and provides them with information on the entangled pair. Alice will claim that the entangled pair still behave locally (the collapse of her wavefunction will not violate her locality), though this locality will be different than the one observed by the deterministic observer.

Alice might guess that Bob sees the entangled pair completely deterministically. She might reason that Bob sits in the UV modes that Alice doesn’t have access to, which she imagines responsible for the state of the entangled pair. But Bob likely does not have access to the IR modes of the entangled pair, where Alice and Carl sit, and views {\em them} as responsible for the state of the entangled pair that he measures. As with Alice, the collapse of Bob’s wavefunction doesn’t violate Bob’s locality, though this locality might be different than the one Alice sees. Ultimately, the physics of the entangled pair is determined by {\em all} the modes of the classical theory. Because Alice, Bob, and Carl do not have access to some very large subset of these modes, they each see a QFT over the very many modes they {\em do} have access too.\\

%---------------------------------------------------------------------------------------------------------------------------------------

\subsubsection{Wigner's Friend}
\label{subsec: wigner}

We’d like to stress the importance 
of associating a mode with Wigner, $\Lambda_W$, Wigner’s friend, $\Lambda_F$, and the system (field) 
that Wigner’s friend is studying, $\Lambda_S$, and placing the three somewhere in an RG hierarchy. We 
will see that where Wigner is situated, and what information he has about the modes 
between $(\Lambda_F,\Lambda_S)$, will determine whether Wigner views the system and/or the friend 
as ``classical" or ``quantum mechanical", and whether Wigner will experience the same locality as his friend. What does it mean for Wigner to observe a ``quantum mechanical” friend? After all, from the friend’s perspective, the system at $\Lambda_S$ is quantum mechanical, and decoherence occurs by the time you get to systems of order $\Lambda_F$. But Wigner might not have access to the modes between $(\Lambda_F,\Lambda_S)$. If he is able to interact with his friend but not with the higher energy modes that have led to the decoherence of his friend, he will just see his friend behaving stochastically, their motion determined by modes which Wigner {\em in principle} (courtesy of P2) cannot interact with.

So let us situate the friend and system (but not yet Wigner) in the following hierarchy:
\be
\Lambda’\ll\Lambda_1\ll\Lambda_F < \Lambda_S\ll\Lambda_2\ll\Lambda~.
\ee
We assume that, in studying the system, Wigner’s friend doesn’t have access to modes 
between $(\Lambda_S,\Lambda)$, integrates them out, and applies RG based on the 
discussion in section \ref{sec: toy model}. So Wigner's friend is looking at a field with a Gaussian fixed point and fine tunes so that the renormalized mass of the system scales as $\mu_S(\xi)\approx\Lambda’\xi=\frac{\Lambda\Lambda’}{\Lambda_S}$. Wigner’s friend will see a QFT for the system they are studying and will see objects of the order of themselves as classical, because of the decoherence that has occurred between modes $(\Lambda_F,\Lambda_S)$.

Wigner might well be an deterministic observer that has information on all modes 
between $(\Lambda’,\Lambda)$ and does not have to integrate anything out. 
Wigner would then see a classical friend and system (both would just be fields 
subject to classical statistical mechanics). But say that Wigner is ignorant of some 
of these modes. When would Wigner see a ``classical" friend? We presume that 
this occurs when Wigner agrees with his friend that the fixed point sets the renormalized 
mass of the system to $\frac{\Lambda\Lambda’}{\Lambda_S}$. To approximate his 
friend as a classical field, Wigner would then follow the steps of section \ref{subsec: classical limit}. For example, 
if Wigner were located at $\Lambda_W\approx\Lambda_F$, he might define a Planck's constant based on $\Lambda_S$;
integrate out additional modes up through $\Lambda_F$; define an effective Planck's constant; and then take the saddle point approximation.
We assume that Wigner was able to see a classical friend because he had information 
on the fixed point at $\Lambda_S$.

The above argument makes sense for $\Lambda_W\approx\Lambda_F$: Wigner has 
about the same information as his friend. If $\Lambda_W\approx\Lambda_1$, the situation 
becomes less clear. It might still be possible for Wigner to approximate his friend based on a fixed 
point of the system his friend is studying. But Wigner is now far from both his friend and the system at $\Lambda_S$ and might only be able to interact with the former. He’d then integrate out modes between $(\Lambda_F,\Lambda)$, set the fixed point at $\mu_F(\xi)\approx\frac{\Lambda\Lambda’}{\Lambda_F}$, and see a quantum mechanical friend. The friend might object: ``decoherence has occurred between me and my external environment!”. Wigner’s response: ``I cannot take a measurement of your external environment. All I see is you behaving stochastically based on modes that I never will have access too”.

The role of friend and system are reversed, for Wigner, if we let $\Lambda_W\approx\Lambda_2$. Wigner now has access to high energy modes but not low energy ones where, relative to Wigner, both the friend and system sit: to Wigner, they are both large scale fields at least the size of his observable universe. Say that Wigner can interact with his friend but with no lower energy modes (at, say, $\Lambda_1$). He’d model his friend by using IRG to integrate out modes between $(\Lambda’,\Lambda_F)$ and see his friend as quantum mechanical: he’d see his friend as a large scale field, interacting with even larger scale modes that Wigner cannot himself measure. If decoherence can then occur over the modes between $(\Lambda_F,\Lambda_S)$, Wigner would see a quantum friend and a classical system! But it’s also possible that Wigner doesn’t have access to his friend’s modes at all. He would then not see his friend, integrate out modes between $(\Lambda’,\Lambda_S)$, and see a quantum mechanical system at $\Lambda_S$. \\

%---------------------------------------------------------------------------------------------------------------------------------------

\subsubsection{RG Invariants}
\label{subsec: RG invariants}

Other than being anti-psi, does RG ontology follow the other postulates of a ``Copenhagenish" quantum mechanics\cite{Schmid:2025ggl}: completeness, universality, and observers observe? The answer seems to formally be ``no” for completeness and universality. After all, we assume that QFT emerges from an underlying classical field in the canonical ensemble. We’d say that deterministic observers who have not integrated out any modes see as complete a picture of the field as is possible: one still subject to the ravages of statistical mechanics, but with all modes accounted for. By definition, the QFTs of non-deterministic observers are missing vast swaths of the RG hierarchy. Let’s ask a more conditional question: what can we say about universality if we confine ourselves to QFTs of non-deterministic observers?

To help us, we define \emph{RG invariants}: quantities that don’t change on application or RG or IRG. For example, this paper deals mostly with non-interactive fields where there is no wavefunction renormalization. So the expressions for Planck’s constant for the UV field, eq.~(\ref{eq: Alice’s planck constant}), is the same as the expression for the IR field, eq.~(\ref{eq: IR hbar and c}), but for switching all $-$ subscripts to $+$ subscript. Thus $\hbar$ is an RG invariant if we only deal with non-interacting scalar fields, but we expect $\hbar$ to \emph{not} be an RG invariant when it starts being renormalized with interactions. 

So perhaps it can be helpful to look for RG invariants when studying universality between different QFT perspectives in an RG hierarchy. What if we confine ourselves to non-deterministic observers who see the \emph{same} QFTs? In section \ref{subsection: Alice's renormalization group}, we state criteria for determining if two applications of the renormalization group lead to the same QFT. By these criteria, we imagine that, yes, for observers who see, say, a UV QFT with the propagator in eq.~(\ref{eq: Feynmann propagator}), things like unitary evolution and locality will be universal. This still leaves unanswered questions like: should states be objectively defined, like in relational QM, or subjectively defined, like in QBism\cite{Fuchs:2016qml}? Perhaps the former would be more plausible if $\psi$ were an RG invariant, but this is beyond the scope of this paper.

Finaly, we imagine an RG ontology with a relativist interpretation of observers observe. All observers, deterministic or not, see definite outcomes, and these are separate from outcomes seen by other observers. So in section \ref{subsection: the measurement problem}, Alice and Carl (who are opposite side of their galaxy) will agree on the eigenstate of two entangled electrons after measurement, but this will in principle be different than the eigenstate observed by Bob when he takes a measurement. Bob lives deep in the UV modes Alice and Carl have integrated out, so it’s assumed that no one, other than a hypothetical deterministic observer, will be able to compare what Alice and Bob measure.

%-----------------------------------------------------------------------------------------------------------------------------------------------------------------
\section{The Non-Interacting Scalar Field Toy Model}
\label{sec: toy model}

We present our toy model for an emergent UV quantum field theory. The goal of this section is twofold. First, it should show how a QFT emerges in the RG ontology interpretation. It should also develop tools we’ll need for studying other observers in the RG hierarchy. With the UV field at hand, we will then move on to section \ref{sec: RG}, where we consider other observers in the hierarchy, as well as the classical limit of the UV field.

%----------------------------------------------------------------------------------------------------------------------------------------
\subsubsection{Classical Physics}
\label{sec: classical physics}

We set the scene with the Lagrange density we will use for the non-interacting scalar field toy model:
\be
\mathcal{L}=\frac 1 2 \gamma_0(\dot{\phi}^2-c_0^2\partial^i \phi\partial_i \phi) - \frac 1 2 \mu_0^2 c_0^4 \phi^2~.
\label{eq: lagrange density}
\ee
The coefficient $\mu_0$ will be trivially renormalized in the non-interacting theory. In a theory with interactions, $\gamma_0$ would also be renormalized and then absorbed into a rescaling of the field. In our model, $\gamma_0$ will also be part of a renormalized Planck’s constant and speed of light. Anticipating the renormalization of the speed of light, we call it $c_0$ and do not set it equal to 1 until we can argue that its renormalization can be nevertheless ignored when dealing with the non-interacting scalar theory. We omit the addition of a constant $\Omega_0$ to eq.~(\ref{eq: lagrange density}), though note in passing that it would be interesting to study how the renormalization of vacuum energy density is framed in an RG ontology interpretation.

Let’s stress that we are working in three dimensional space, so that the Hamiltonian is $H=\int\mathrm{d}^3x\mathcal{H}$. This is in contrast to the correspondence between a quantum field theory in four dimensional spacetime and a classical field theory in four dimensional space. We do not write $H=\int\mathrm{d}^4x\mathcal{H}$ because we think that this would not be physical for a deterministic observer (see section \ref{subsec: QFT_postulates}). Instead, we view both the QFT and the 4D spatial classical theory as emerging from a three (spatial) dimensional classical field theory placed in an RG hierarchy. To get the time integral, we use Hamilton-Jacobi formalism. We assume that the Hamiltonian is evaluated at some final field configuration, $\phi$, which occurs at some final time t (where the initial field configuration, assume fixed, occurs at time $t_1=0$). We write the Hamiltonian as $H(t)$ (omitting notation for the functional dependence on the field), and then write
\be
-H(t)=L(t)-\int\mathrm{d}^3x~\pi\dot{\phi}=\partial_t \left(S(t)-W_0(t)\right)~,
\label{eq: HJ}
\ee
where we recognize the second expression on the right as the Hamilton characteristic function in the case that there is no explicit time dependence in the Hamiltonian:
\be
W_0(t)=\int^t_0 \mathrm{d}^4x ~\pi\dot{\phi}~.
\ee

We will renormalize in k-space, so we first Fourier transform the action and principle function. We use the notation $\phi(x)=\int\frac{\mathrm{d}^4k}{(2\pi)^4}e^{i k\cdot x}\phi(k)$ and emphasize that
the dot product is Euclidean. The action then becomes, for the Lagrange density of eq.~(\ref{eq: lagrange density}),
\bea
S(t)&=&\int\frac{\mathrm{d}^4k_1\mathrm{d^4}k_2}{(2\pi)^8}\mathrm{d}^3\vec{x}\int^{t}_{0}\mathrm{d}x_0~e^{i x\cdot(k_1+k_2)}\phi(k_1)\phi(k_2)\left[\gamma_0(-k_{10}k_{20}-c_0^2\vec{k}_1\cdot\vec{k}_2)-\mu_0^2c_0^4\right]\notag \\
&=&
-\frac{1}{2}\int\frac{\mathrm{d}^4k_1\mathrm{d}k_{20}}{(2\pi)^5}\phi(k_1)\phi(-\vec{k}_1,k_{20})\frac{\gamma_0(k_{10}k_{20}+c_0^2\vec{k}_1^2)+\mu_0^2c_0^4}{i(k_{10}+k_{20})}\left[e^{i t (k_{10}+k_{20})}-1\right]~,
\eea
where we've performed the integrals over $\vec x$, $\vec k_2$, and $x_0$ in the second line. We then do the $k_{20}$ integral by adding the small momentum $-i\epsilon$ to $k_{10}$
and performing the contour integration in the upper half plane. The result is
\be
S(t)=-\frac 1 2 \left[e^{-\epsilon t}-1\right]\int\frac{\mathrm{d}^4k}{(2\pi)^4}\phi(k)\phi(-\vec k,k_0+i\epsilon)\left[\gamma_0(-k_0^2+i\epsilon k_0+c_0^2\vec{k}^2)+\mu_0^2c_0^4\right]~.
\label{eq: action fourier}
\ee
We repeat these steps for the Hamilton characteristic function. Using eq.~(\ref{eq: lagrange density}), the generalized momentum is $\pi=\frac{\partial\mathcal{L}}{\partial\dot{\phi}}=\gamma_0\dot{\phi}$,
and we have
\be
W_0(t)=\gamma_0\int^{t}_{0}\mathrm{d}^4x~\dot{\phi}^2 =
\gamma_0\left[e^{-\epsilon t}-1\right]\int\frac{\mathrm{d}^4k}{(2\pi)^4}\phi(k)\phi(-\vec k,-k_0+i\epsilon)(k_0^2-i\epsilon k_0)~.
\label{eq: principle function fourier}
\ee
We can now substitute eq.~(\ref{eq: principle function fourier}) and eq.~(\ref{eq: action fourier}) into the expression for the Hamiltonian in eq.~(\ref{eq: HJ}):
\be
-H(t)=\frac{\epsilon}{2}e^{-\epsilon t}\int\frac{\mathrm{d}^4k}{(2\pi)^4}\phi(k)\phi(-\vec k,-k_0+i\epsilon)\left[\gamma_0(k_0^2-i\epsilon k_0+c^2\vec{k}^2)+\mu_0^2c^4\right]~.
\label{eq:nonrel Hamiltonian}
\ee

What is $\epsilon$? We got it in eq.~(\ref{eq: action fourier}) after doing a contour integration in $k_{10}$. Similar factors emerge when generally studying the correspondence between a classical partition function and a quantum path integral. We’d get it, for example, when comparing the classical 1D Ising model partition function to a quantum partition function with one degree of freedom\cite{Shankar_2017}. There, the coefficients in the free energy of the classical Hamiltonian can be expressed in terms of a small parameter $\epsilon$ and the Lagrangian coefficients. We will instead try absorbing our $\epsilon$ into the renormalized Planck constant. We ultimately view $\epsilon$ as an artifact of an attempt to renormalize four dimensional spacetime starting with a classical theory in a three dimensional space.

%---------------------------------------------------------------------------------------------------------------------------------------

\subsubsection{Postulates 1 and 2}
\label{subsection: postulates 1 and 2}

We situate our model in the RG hierarchy
\be
\Lambda’\ll\Lambda_1\ll\Lambda~.
\label{eq: RG hierarchy, UV and IR theories}
\ee
The idea is to pass to the non-deterministic observer by integrating out modes up through $\Lambda_1$. We note that the modes in eq.~(\ref{eq: RG hierarchy, UV and IR theories}) are magnitudes in the four-dimensional k-space. We are working with non-interacting scalar theory, so in addition to the lack of mode mixing in eq.~(\ref{eq:mode seperation}), we can do the functional integral over $\pi$ in the partition function and absorb the result into the integration measure. So the partition function is
\be
Z_{\mp}=\int\mathcal{D}\phi^{\mp}e^{-\beta H[\phi^{\mp}]}~,
\ee
with the functional measure
\be
\mathcal{D}\phi^{\mp}=\left[\frac{1}{A} \sqrt{\frac{-2\pi\gamma_0}{\beta \Delta V}}\right]^N \left(\int\mathcal{D}\phi^{\pm}e^{-\beta H[\phi^{\pm}]}\right)\prod_i \mathrm{d}\phi^{\mp}_i~.
\label{eq: functional measure}
\ee
We've includes the functional integral over $\phi_{\pm}$ modes that were originally part of the effective action. The functional measure includes the very small spatial volume $\Delta V$. It also includes the constant $A$, added to make the partition function dimensionless (it should have the same dimension as $\phi\pi$). We do not attempt to calculate the overall constant in the partition function or come up with an expression for $A$, though we imagine that the latter is proportional to the emergent Planck’s constant. This is because the factors of Planck’s constant that appear in, say, the calculation of the classical partition function for an ideal gas can be derived with many body quantum mechanics, and the latter can be expressed as a field theory. Throughout this paper, we set the Boltzmann constant to 1, so that $\beta^{-1}=T$.
  
We then rescale the momenta and fields according to
\be
k'_0=\xi_{\mp}k_0~~~,~~~\phi^{'\mp}(k')=\gamma_{\mp}\xi_{\mp}^{-3}\phi^{\mp}(k)~.
\label{eq: general rescaling}
\ee
Note that the scaling factor $\xi_{\mp}$ has yet to be defined. The notation for gamma is just to remind us that it would be renormalized if we had interactions: for most of the present paper, $\gamma_{\mp}=\gamma_0$. But we include it to illustrate a hypothetical renormalization scheme for $\hbar$ and $c$ that we will discuss in \ref{subsection: hbar, c, and postulate 3}. The Hamiltonian becomes
\be
-H[\phi^{\mp}]=\frac{\epsilon e^{-\epsilon t}}{2\gamma_{\mp}}\int^{\Lambda_{\mp}}_{\Lambda'_{\mp}}\frac{\mathrm{d}^4k}{(2\pi)^4}\phi^{\mp}(k)\phi^{\mp}(-\vec k,-k_0+i\epsilon\xi_{\mp})
\left[k_0^2-i\epsilon k_0\xi_{\mp}+c^2_0\vec{k}^2+\mu_{\mp}^2 c_0^4\gamma_{\mp}\right]~,
\label{eq:renormalized Hamiltonian}
\ee
where the renormalized mass is
\be
\mu_{\mp}=\mu_0\xi_{\mp}\gamma_{\mp}^{-1}~.
\label{eq: renormalized mass}
\ee
Being situated in the RG hierarchy, the original bounds before rescaling satisfied either $\Lambda’\ll\Lambda_1$ or $\Lambda_1\ll\Lambda$. As a result, the rescaled bounds will all have $\Lambda’_{\mp}\ll\Lambda_{\mp}$. To satisfy the cutoff rescaling criteria of section \ref{subsec: QFT_postulates}, it is left to choose $\xi_{\mp}$ so that one of the rescaled bounds equals one of the original bounds, either $\Lambda’$ or $\Lambda$. We will see a couple of ways that this can be done here and in section \ref{sec: RG}.

We now specialize to the UV field, saving the IR field for section \ref{subsection: IR QFT}. We set $\xi_{-}=\frac{\Lambda}{\Lambda_1}$,
which sets the rescaled bounds to
\be
\Lambda'_{-}=\Lambda'\xi_{-}~~~,~~~\Lambda_{-}=\Lambda~.
\label{eq: Alice bounds}
\ee
Note that applying postulate 2 {\em before} allowing RG flow prevents
the lower cutoff from flowing to the upper cutoff in eq.~(\ref{eq: Alice bounds}). This means, for example, that while $\xi_{-}$ will increase,
it will always be true that $\xi_{-}\ll\frac{\Lambda}{\Lambda'}$.

%---------------------------------------------------------------------------------------------------------------------------------------

\subsubsection{$\hbar$, c, and Postulate 3}
\label{subsection: hbar, c, and postulate 3}

We now define a Planck’s constant for the UV theory,
\be
\hbar_{-} \equiv \frac{T\gamma_{-}}{\epsilon}~,
\label{eq: Alice’s planck constant}
\ee
and a renormalized speed of light,
\be
c_{-}\equiv c_0\gamma_{-}^{\frac{1}{4}}~,
\label{eq: renormalized speed of light}
\ee
through which the Hamiltonian becomes
\be
-\beta H[\phi^{-}]=\frac{e^{-\epsilon t}}{2\hbar_{-}}\int^{\Lambda_{-}}_{\Lambda'_{-}}\frac{\mathrm{d}^4k}{(2\pi)^4}\phi^{-}(k)\phi^{-}(-\vec k,-k_0+i\epsilon\xi_{-})
\left[k_0^2-i\epsilon k_0\xi_{-}+c^2_0\vec{k}^2+\mu_{-}^2 c_{-}^4\right]~.
\label{eq: renormalized H with hbar and c}
\ee
Note that eq.~(\ref{eq: renormalized H with hbar and c}) has both a $c_0$ and a $c_{-}$. We do not claim that eqs.~(\ref{eq: Alice’s planck constant},~\ref{eq: renormalized speed of light}) are \emph{the} Planck’s constant and speed of light, only that they are sufficient for non-interacting scalar QFT. We refrain from calling eq.~(\ref{eq: Alice’s planck constant}) an ``effective” Planck’s constant and reserve this phrase for section \ref{subsec: classical limit}, where we discuss the classical limit to Alice’s field.

We included $\gamma_{-}$ in eqs.~(\ref{eq: Alice’s planck constant},~\ref{eq: renormalized speed of light}) because we speculate that, in a theory with interactions, wavefunction renormalization might ultimately lead to the renormalization of $\hbar_{-}$ and $c_{-}$. For example, in $\phi^4$ scalar QFT, $\gamma_{-}$ would be, to two loops,
\be
1-\gamma_{-}\propto\lambda_0^2\ln\frac{\Lambda}{\Lambda_1}~,
\label{eq: renormalized gamma with phi^4}
\ee
where $\lambda_0$ is the bare coupling. \emph{If} eq.~(\ref{eq: renormalized gamma with phi^4}) applied in eq.~(\ref{eq: Alice’s planck constant}) and eq.~(\ref{eq: renormalized speed of light}), we’d satisfy the flow criteria in the introduction: as we integrate out more modes in the UV theory, $\hbar_{-}$ and $c_{-}$ would decrease. This is the reason for our choice of scaling in eq.~(\ref{eq: general rescaling}). Please note: we do not give the derivation of eq.~(\ref{eq: renormalized gamma with phi^4}) in the RG ontology model, nor do we give a method here for consistently renormalizing $\hbar_{-}$ and $c_{-}$ when interactions are present. The $\gamma_{-}$ in eq.~(\ref{eq: Alice’s planck constant}) and eq.~(\ref{eq: renormalized speed of light}) is only suggestive in the context of our toy model. 

The toy model should be for a non-interacting QFT in the vacuum, so why do we still have a temperature in eq.~(\ref{eq: Alice’s planck constant})? The previous paragraph not withstanding, $\gamma_0$ in \emph{not} renormalized in our non-interacting scalar theory, and so neither is Planck’s constant. We assume that we can set non-renormalized quantities to 1. So if $\hbar_{-}=1$, eq.~(\ref{eq: Alice’s planck constant}) tells us that $T\approx\epsilon$. We assume that the non-interactive theory is at this temperature.  If we were looking at $\phi^4$ theory, we’d instead imagine that the bare mass would satisfy $\mu_0\approx T-T_c$ in the mean field approximation, so that $\hbar_{-}\propto T_c$. This would be for the vacuum theory. The non-deterministic observe would see a field at 0K while the deterministic observer sees one at $T_c$. But we wait for the fuller development of the story with the interactive theory.

Let us go to the Gaussian fixed point with the UV toy model. Just like $\hbar_{-}$, because $\gamma_{-}$ doesn’t flow as we integrate modes, neither does $c_{-}$, so we should be able to set $c_0=c_{-}=1$. By ``Gaussian fixed point”, we mean that the renormalized mass should flow to the lower rescaled cutoff. This is accomplished by fine tuning the bare mass to hit the bare lower cutoff,
\be
\mu_0\approx\Lambda’\rightarrow \mu_{-}\approx\Lambda’_{-}=\frac{\Lambda\Lambda’}{\Lambda_1}~.
\ee

The correlation length is $l_0=\frac{\sqrt{\gamma_0}}{\mu_0}$ and we would like it and the renormalized correlation length, $l=l_0/\xi_{-}$, to be divergent. Comparing to the time \emph{t} of some final field configuration, introduced in eq.~(\ref{eq: HJ}), this would mean that we should have
\be
\mu_0 t\ll1~~~\mathrm{and}~~~\mu_{-} t\ll1~.
\label{eq: time limit}
\ee
The first condition of eq.~(\ref{eq: time limit}), involving the bare correlation length, demands that $t^{-1}\gg\Lambda’$. We make the choice
\be
t\approx\frac{1}{\Lambda_1}~,
\ee
so that $t^{-1}$ is of order of the field that Alice is studying. Note that we can now ignore the factor $e^{-\epsilon t}$ in eq.~(\ref{eq: renormalized H with hbar and c}) because we assume at least that $\epsilon\approx \Lambda’$, and therefore $\epsilon t\approx \frac{\Lambda’}{\Lambda_1}\ll1$. The condition $\mu_{-}t\ll1$ becomes a restriction on the bounds beyond eq.~(\ref{eq: RG hierarchy, UV and IR theories}),
\be
\frac{\Lambda}{\Lambda_1}\ll\frac{\Lambda_1}{\Lambda'}~.
\ee
We’ll see the limits in eq.~(\ref{eq: time limit}) play a role in the propagator discussion below.

%---------------------------------------------------------------------------------------------------------------------------------------

\subsubsection{Propagator}
\label{subsection: propagator}

The claim in the introduction is that the RG ontology model, after the application of the postulates, is, to the non-deterministic observer, a QFT. We check this by applying the postulates to the Green’s function and checking that it has the expected causal structure for a non-interacting scalar QFT: we would like to be able to approximate it as the modified Bessel function of the 2nd kind, because we want to reproduce the behavior of the Feynmann propagator. We take the standard route to the Green’s function: add a source term $J\phi$ to eq.~(\ref{eq: lagrange density}) and then let $Z_{J=0}=1$. Once we renormalize, the propagator becomes
\be
\Delta_{\mp}(r)=-\frac{e^{-\epsilon t}}{\gamma_{\mp}}\int^{\Lambda_\mp}_{\Lambda'_{\mp}}\frac{\mathrm{d}^4k}{(2\pi)^4}\frac{e^{-k\cdot r}}{k_0^2-i\epsilon k_0\xi_{\mp}+c^2_0\vec{k}^2+\mu_{\mp}^2c_{\mp}^4}.
\label{eq: nonrenormalized GF}
\ee
We note that one of the factors of $\gamma_{\mp}$ in the denominator comes from us wanting an overall factor of $\gamma_{\mp}$ in the Hamiltonian that gets absorbed into Planck’s constant. 

We work with the UV version of eq.~(\ref{eq: nonrenormalized GF}), set the speed of light and $\gamma_{-}$ to 1, and nix the factor of $e^{-\epsilon t}$. We will also ignore the $i\epsilon k_0\xi_{-}$ term in the denominator, notwithstanding the fact that, if the cutoffs could be removed, it would be used to evaluate $\Delta_{-}$ via contour integration. Instead, we apply P2 and P3 to
\be
\Delta_{-}(r)=-\int^{\Lambda_{-}}_{\Lambda’_{-}}\frac{\mathrm{d}^4k}{(2\pi)^4}\frac{e^{-k\cdot r}}{k^2+\mu_{-}^2}~,
\ee
where we $k$ and $r$ are magnitude in Euclidean space. Performing the angular integration, we get
\be
\Delta_{-}(r)=-\frac{1}{(2\pi)^2 r}\int^{\Lambda_{-}}_{\Lambda’_{-}}\frac{\mathrm{d}k~k^2}{k^2+\mu_{-}^2}J_{1}(kr)~.
\label{eq: propagator angular integral}
\ee

We now change variables to $k=\mu_{-}\tan{\theta}$ and evaluate the integration bounds with the postulates in mind. P3 sets $\Lambda’_{-}=\mu_{-}$ so that the new lower bound becomes $\pi/4$. For the new upper bound, we have $\tan\theta=\frac{\Lambda_{-}}{\mu_{-}}\approx\frac{\Lambda_1}{\Lambda’}$, so that the new upper bound is $\pi/2$ with application of P2. We thus have
\be
\Delta_{-}(r)=-\frac{1}{(2\pi)^2}\frac{\mu_{-}}{r}\left[\int^{\pi/2}_0 \tan^2\theta~J_1(\mu_{-}r\tan\theta)\mathrm{d}\theta-\int^{\pi/4}_0 \tan^2\theta~J_1(\mu_{-}r\tan\theta)\mathrm{d}\theta \right]~.
\label{eq: GF variable change}
\ee
The second term of eq.~(\ref{eq: GF variable change}) is much smaller than the first. We can see this by assuming that the argument of the Bessel functions is small like with eq.~(\ref{eq: time limit}), $\mu_{-}r\approx \mu_{-}t\ll1$, and then approximating $J_1(\mu_{-}r\tan\theta)\approx\frac{1}{2}\mu_{-}r\tan\theta$ for the integral in the second term. We get the desired result:
\be
\Delta_{-}(r)=-\frac{1}{(2\pi)^2}\frac{\mu_{-}}{r} K_1(\mu_{-}r)~.
\label{eq: Feynmann propagator}
\ee
%

%---------------------------------------------------------------------------------------------------------------------------------------

\subsubsection{Commutation Relations}
\label{subsection: commutation relations}

Setting $c=1$ and ignoring terms with $\epsilon$ (other than for its presence in Planck’s constant), we have the Hamiltonian
\be
-H[\phi^{-}]\beta=\frac{1}{2\hbar_{-}}\int^{\Lambda_{-}}_{\Lambda’_{-}}\frac{\mathrm{d}^4k}{(2\pi)^4}\phi^{-}(k)\phi^{-}(-k)[k_0^2+\vec{k}^2+\mu_{-}^2]~.
\label{eq: Alice's renormalized H at fixed point}
\ee
At this point, we are tempted to remove the cutoffs and do a Wick rotation. This would give us a direct analog with the action of non-interacting scalar QFT. But we do not provide an argument here on how this would work with, say, a rescaled lower bound in $\Lambda'_{-}$. This process would also not be generalizable to, say, $\phi^4$, because removing cutoffs there leads to a trivial theory. We at least point out that the link between eq.~(\ref{eq: Alice's renormalized H at fixed point}) and the action in QFT would motivate a \emph{derivation} of commutation relations from the partition function in QFT rather than the other way around. 

So we briefly review the derivation of commutation relations, starting from the QFT partition function
\be
Z[J]=\int\mathcal{D}\phi~\mathrm{exp}\left\{-i\left[\frac{1}{2}\phi(-\partial^2+\mu^2)\phi-\phi J\right]\right\}~.
\label{eq: QFT partition function with operatores}
\ee
Setting the sourse to 0 and noting that we are working with the ground state, we also have $Z[J]=\mathrm{exp}\left\{\frac{i}{2} J\Delta J\right\}$. Taking two functional derivative of this and eq.~(\ref{eq: QFT partition function with operatores}), we get 
\be
\Delta(x-x')=\langle i\mathrm{T}\phi(x)\phi(x') \rangle~.
\label{eq: time ordering}
\ee
Plugging the right side of eq.~(\ref{eq: time ordering}) into $(-\partial^2+m^2)\Delta(x-x')=\delta^4(x-x')$, we get the equal time commutation relation between $\phi$ and $\pi$.

We close this section with a discussion of locality. We argued that the speed of light might generally be renormalized. If we can’t ignore this effect, we should not set $c=1$, which would imply that it is not. But it is certainly not renormalized in the non-interacting theory. We had thus set $c=1$ above and found that application of P2 and P3 got us the correct causal structure for the propagator in eq.~(\ref{eq: Feynmann propagator}). Finally, we can go to Minkowski space if we remove the cutoffs in eq.~(\ref{eq: Alice's renormalized H at fixed point}). The local theory we are left with in eq.~(\ref{eq: QFT partition function with operatores}) is to be contrasted with the non-local theory we started with, in eq.~(\ref{eq:nonrel Hamiltonian}).

%-------------------------------------------------------------------------------------------------------------------------------------------------------------

\section{Up and Down the RG Hierarchy}
\label{sec: RG}

In section \ref{sec: toy model}, we studied the emergence of QFT for Alice’s UV field. But in much of section \ref{sec: philosophy}, we stressed that there should be other observers in the RG hierarchy who may or may not see local quantum mechanical phenomena depending on the information (modes) they have access too. These observers might be located in strange places (in, for example, modes that another observer has integrated out) and may or may not see a QFT where other observers in the hierarchy do or do not. The objective of this section is to study these other observers.

%---------------------------------------------------------------------------------------------------------------------------------------

\subsection{The Classical Limit and Wigner's Friend}
\label{subsec: classical limit}

Let’s discuss again Wigner’s friend thought experiment, which we had previously discussed in section \ref{subsec: foundations} but now armed with our work in the non-interacting UV theory. We consider the following hierarchy of modes:
\be
\Lambda’\ll\Lambda_W\ll\Lambda_A<\Lambda_1\ll\Lambda~,
\label{eq: classical limit hierarchy}
\ee
where the observers are Wigner, at $\Lambda_W$, and Alice, at $\Lambda_A$. We imagine that Alice has evidence for a QFT at $\Lambda_1$, which she then models by integrating out modes between $(\Lambda_1,\Lambda)$ and then applying the postulates of section \ref{subsec: QFT_postulates}. Alice’s Planck’s constant and renormalized speed of light are the same as in section \ref{subsection: hbar, c, and postulate 3}:
\be
c_{1'}= c_0\gamma^{\frac{1}{4}}_{1'} ~~~\mathrm{and}~~~\hbar_{1'}\approx T\gamma_{1'}\epsilon^{-1}~.
\label{eq: new notation}
\ee
Eq.~(\ref{eq: new notation}) introduces new notation we'll use throughout section \ref{sec: RG}: if the RG hierarchy has more than one mode between $\Lambda'$ and $\Lambda$, subscripts on renormalized quantities and fields indicate which modes were not integrated out. We write the subscript for the larger mode first, write a prime for $\Lambda'$, and omit a subscript for $\Lambda$. So $\hbar_{1'}$ indicates that RG was used and that the modes $(\Lambda',\Lambda_1)$ weren't integrated out while $\hbar_{1}$ indicates that IRG was used and that we were left with modes between $(\Lambda_1,\Lambda)$.

Alice might then want to model objects around where she is, at $\Lambda_A$, by claiming further ignorance about modes between $(\Lambda_A,\Lambda_1)$. She integrates out the additional modes to get the following Hamiltonian:
\be
-\beta H[\phi^{-}]=\frac{1}{2\hbar_{A'}}\int^{\Lambda_{A'}}_{\Lambda'_{A'}}\frac{\mathrm{d}^4k}{(2\pi)^4}\phi_{A'}(k)\phi_{A'}(-k)
\left[k_0^2+c^2_0\vec{k}^2+\mu_{A'}^2 c_{A'}^4\right]~.
\label{eq: Alice’s effective hamiltonian}
\ee
Note again that $\phi_{A'}$ is for a field defined over the modes $(\Lambda’,\Lambda_A)$. The bounds are $\Lambda_{A'} = \Lambda$ and $\Lambda’_{A'}=\frac{\Lambda’\Lambda}{\Lambda_A}$ with the renormalized mass $m _{A'} = \frac{\mu_0}{\gamma_{A'}}\frac{\Lambda}{\Lambda_A}$. We can then define an effective Planck's constant for Alice and compare to her ``actual" Planck's constant:
\be
\hbar_{\mathrm{eff}}^{\mathrm{Alice}} < \hbar^{\mathrm{Alice}}~,
\label{eq: observer label}
\ee
with $\hbar_{\mathrm{eff}}^{\mathrm{Alice}} = \hbar_{A'}$ and $\hbar^{\mathrm{Alice}}=\hbar_{1'}$, and where we've indicated the observer that the Planck's constants correspond to. We assume eq.~(\ref{eq: observer label}) is true because we assume that both $\gamma_{A'}$ and $\gamma_{1'}$ obey an equation like eq.~(\ref{eq: renormalized gamma with phi^4}), but we don't show this explicitly. With her effective Planck’s constant smaller than her actual Planck’s constant, Alice might be able to use the saddle point approximation. Note that there is a similar expression for the effective renormalized speed of light:
\be
c_{\mathrm{eff}}^{\mathrm{Alice}}< c^{\mathrm{Alice}}~,
\ee
with $c_{\mathrm{eff}}^{\mathrm{Alice}}=c_{A'}$ and $c^{\mathrm{Alice}}=c_{1'}$.

What Wigner sees depends on which modes he has information on. Assume for this discussion that he has access to $\Lambda_A$ but nothing larger. Wigner’s Planck constant is $\hbar_{A'}$, which is Alice’s effective Planck’s constant, but Wigner cannot take the saddle point approximation while Alice can! That $\hbar_{A'} < \hbar_{1'}$ is immaterial to Wigner as he doesn’t know about $\hbar_{1'}$. The speculation is that Alice has more information than Wigner, even though she’s integrated out the same number of modes. She is aware, perhaps from a separate experiment, that there is quantum mechanical behavior (for her) at $\Lambda_1$, and this is enough for her to potentially see deterministic physics at $\Lambda_A$.

The discussion of locality between Wigner and Alice follows the same logic as the discussion of Planck’s constant. While Wigner and Alice see the same renormalized speed of light, $c_{A'}$, Alice is aware of the larger speed $c_{1'}$, presumably through a separate experiment, and defines her locality based on the latter. Both perspectives will be local in their own right courtesy of P2 and P3. But if they were somehow able to compare their relativistic invariants, they’d not agree on them, unless some of those invariants turned out to be RG invariants. We presume that only a deterministic observer (one who hasn't integrated out any modes, which neither Alice nor Bob are) would actually be able to make such a comparison.

%-----------------------------------------------------------------------------------------------------------------

\subsection{Emergent QFT for an IR Classical Field Theory}
\label{subsection: IR QFT}

We study the potential emergence of a QFT when an observer integrates out the IR modes up through $\Lambda_1$ in the hierarchy of eq.~(\ref{eq: RG hierarchy, UV and IR theories}). If such an observer sees an emergent QFT, this would have nothing to do with any QFT they’d \emph{separately} observe for a UV field. Instead, it comes from their ignorance of large scale modes. In principle, we can imagine a similar formula for Planck’s constant and the renormalized speed of light as in section \ref{subsection: hbar, c, and postulate 3}:
\be
\hbar_{+}=T\gamma_{+}\epsilon^{-1}~~~,~~~c_{+}=c_0\gamma_{+}^{\frac{1}{4}}~.
\label{eq: IR hbar and c}
\ee
But it’s not clear, without an expression for the flow of $\gamma_{+}$, whether we should scale the field as we did in eq.~(\ref{eq: general rescaling}). We do not do the two loop calculation here for the IR field, and so the ultimate form of $\hbar_{+}$ and $c_{+}$ will need to wait for the interactive theory. For the rest of this section we again note that $\gamma_{+}$ is not renormalized in the non-interactive theory and use this to set $h_{+}=1$ and $c_{+}=1$.

P2, along with the cutoff rescaling criteria, seems to have increased the number of fixed points we can study for a non-interacting scalar field, determined by both the fine tuning of $\mu_0$ and the scaling factor $\xi_{+}$. In general, the rescaled bounds for the IR field are $\Lambda’_{+}=\Lambda_1\xi_{+}$ and $\Lambda_{+}=\Lambda$ while the renormalized mass is $\mu_+=\mu_0\xi_{+}$. So the fixed point is ``massless” if $\mu_0\approx\Lambda’$, Gaussian if $\mu_0\approx\Lambda_1$, and the fixed point at infinity if $\mu_0\approx\Lambda$. 

We briefly consider each of these possibilities. None of the fields we’ll study in this section will be an exact match to the QFT of section \ref{sec: toy model}, but we don’t necessarily expect this to have been the case. The point is that both perspective on the same field might live somewhere out there, but be ignorant of each other, because of their place in an RG hierarchy.\\

%---------------------------------------------------------------------------------------------------------------------------------------

\paragraph{Gaussian fixed point.} We set the scaling factor to 
\be
\xi_{+}=\frac{\Lambda’}{\Lambda_1}~,
\label{eq: IR standard scaling factor}
\ee
which fulfills the cutoff rescaling criteria with 
\be
\Lambda’_{+}=\Lambda’~~~,~~~\Lambda_{+}=\frac{\Lambda\Lambda’}{\Lambda_1}~.
\ee
Whether we can consistently fine tune with $\mu_0\approx\Lambda_1$ in the interactive IR theory remains unclear. This also makes the bare mass larger than $\Lambda’$ but the renormalized mass equal to $\Lambda_{+}’$, which is not what we had in the non-interacting UV theory. If we wanted to satisfy P3 through the IR version of eq.~(\ref{eq: time limit}), we’d have to study $t^{-1}\gg\Lambda_1$. We might choose, say, $t^{-1}\approx\frac{\Lambda_1^2}{\Lambda’}$, but also set the bounds so that $t^{-1}\ll\Lambda$. On the other hand, we’d get the same expression for the propagator as eq.~(\ref{eq: Feynmann propagator}).\\

%---------------------------------------------------------------------------------------------------------------------------------------
\paragraph{Massless fixed point.} We finetune with $\mu_0\approx\Lambda’$ and again  have the scaling factor in eq.~(\ref{eq: IR standard scaling factor}). The renormalized mass now scales as
\be
\mu_+\approx\Lambda'\frac{\Lambda’}{\Lambda_1}\ll\Lambda'=\Lambda’_{+}~.
\label{eq:mu_+ for gaussian fixed point}
\ee
Again the relation between the bare mass and cutoffs is not the same as that between the renormalized mass and cutoffs. Let’s see how eq.~(\ref{eq:mu_+ for gaussian fixed point}) affects the propagator. We go to evaluate
\be
\Delta_+ (r) = \frac{-1}{(2\pi)^2}\frac{1}{r}\int_{\Lambda’_+}^{\Lambda_+}\frac{k^2~\mathrm{d}k}{k^2+\mu_+^2}J_1 (kr).
\label{eq: propagator for gaussian fixed point}
\ee
This time, we cannot resort to the variable change below eq.~(\ref{eq: propagator angular integral}). Like with the UV field, we make the choice $t^{-1}\approx\Lambda_1$, so to get a large correlation length, with which we have a large rescaled correlation length, or 
\be
\mu_+t\approx\left(\frac{\Lambda'}{\Lambda_1}\right)^2\ll1~.
\label{eq: IR gaussian fixed point trivial bounds condition}
\ee
The leading approximation to eq.~(\ref{eq: propagator for gaussian fixed point}) is then
\be
\Delta_+ =\frac{-1}{(2\pi)^2}\frac{J_0 (\Lambda’_+ r)-J_0 (\Lambda_+ r)}{r^2}~.
\label{eq: IR propagator}
\ee
Which is a far cry from eq.~(\ref{eq: Feynmann propagator}).\\

%---------------------------------------------------------------------------------------------------------------------------------------

\paragraph{Fixed Point at Infinity.} With the Gaussian fixed point no longer determined by a flow towards ``0”, we don’t see why we shouldn’t give more attention to fields that flow to the upper cutoff rather than the lower one. Indeed, in this section we’ll see behavior of the propagator similar to the UV field. With the choice of scaling in eq.~(\ref{eq: IR standard scaling factor}), we fine tune the bare mass to $\mu_0\approx\Lambda$, which hits the fixed point at infinity:
\be
\mu_0\approx\Lambda\rightarrow\mu_+\approx\frac{\Lambda\Lambda’}{\Lambda_1}=\Lambda_+~.
\ee
A drawback is that we no longer satisfy postulate 3: if we impose the condition $\mu_0 t\ll1$, we are forced to choose the time $t^{-1}\gg\Lambda$, which seems unreasonable given that the original field was defined over a range smaller than this.
So for now, we consider limits of a small correlation length, were
\be
\mu_0 t\gg1~~~\mathrm{and}~~~\mu_+ t\gg1~.
\ee
This at least synchronizes a choice of time with that made for the UV field, $t^{-1}\approx\Lambda_1$, though the condition on the bounds has reversed:
\be
\frac{\Lambda}{\Lambda_1}\gg\frac{\Lambda_1}{\Lambda’}~.
\ee
What about the propagator? From eq.~(\ref{eq: propagator for gaussian fixed point}), we set $k=\mu_+ \tan\theta$ like we did in section \ref{subsection: propagator}. This time,
\be
\Delta_+ (r)=-\frac{1}{(2\pi)^2}\frac{\mu_+}{r}\int^{\frac{\pi}{4}}_0 \tan^2\theta~J_1(\mu_+ r \tan\theta)\mathrm{d}\theta~.
\ee
With $\mu_+ r \gg 1$, we again get a Bessel function of the 2nd kind to leading order,
\be
\Delta_+(r)=-\frac{1}{(2\pi)^2}\frac{K_1 (\mu_+ r)}{2r^2}~.\\
\label{eq: IR propagator at infinity}
\ee
%

%---------------------------------------------------------------------------------------------------------------------------------------

\paragraph{Fixed Point at Infinity and Postulate 3.} Could we have a fixed point at infinity and also satisfy postulate 3? We keep the finetuning $\mu_0\approx\Lambda$, but we now choose the same scaling that we did for the UV field,
\be
\xi_{+}=\frac{\Lambda}{\Lambda_1}~,
\ee
so that the rescaled bounds become
\be
\Lambda’_+ = \Lambda~~~\mathrm{and}~~~\Lambda_{+}=\frac{\Lambda^2}{\Lambda_1}~.
\ee
What was once the high energy cutoff has turned into the low energy cutoff! Whether we can do this consistently with interactions isn’t clear, but we’ve satisfied the cutoff rescaling through the lower rescaled cutoff (and the limits between the two rescaled cutoffs). The choice $t^{-1}\gg\Lambda$ now seems a little more palatable, at least to the non-deterministic observer. The propagator now takes the form of the 2nd term in eq.~(\ref{eq: GF variable change}) and is now independent of $r$:
\be
\Delta_+ (r) = -\frac{1-\log 2}{(2\pi)^2}\frac{\mu_+^2}{4}~.
\ee
%

%---------------------------------------------------------------------------------------------------------------------------------------

\subsection{The Passive and Active Pictures}
\label{subsection: Active/Passive}

%---------------------------------------------------------------------------------------------------------------------------------------

\paragraph{Passive Picture.} How do we compare the results of sections \ref{sec: toy model} and \ref{subsection: IR QFT}? In the passive picture, Alice’s UV field is the same object as Bob’s IR field. We seem to have two options for the passive picture: either assume that Alice’s fine tuning also applies to Bob or assume that they separately fine tune their respective bare parameters. The second option, where Alice and Bob each separately fine tune, is more in line with the philosophy of this paper, because we tend to assume that Alice and Bob have no access to the information that the other has (information on the low energy modes in Alice’s case, the high energy modes in Bob’s case). So if, looking at the same field, Alice sees the Gaussian fixed point while Bob sees the fixed point at infinity, we'd have the fine tunings
\be
\mu_{0}^{\mathrm{Alice}}\approx \Lambda’~~~\mathrm{and}~~~\mu_{0}^{\mathrm{Bob}}\approx\Lambda~.
\label{eq: fine tuning 1, passive picture}
\ee

Alice and Bob would see different physics for the same field. Assuming the fine tuning in eq.~(\ref{eq: fine tuning 1, passive picture}), we could compare the resulting propagators in eq.~(\ref{eq: Feynmann propagator}) for Alice vs. eq.~(\ref{eq: IR propagator at infinity}) for Bob. We’d assume that Alice’s and Bob’s $\hbar$ and $c$ would also be different because of the different wavefunction renormalization in $\gamma_{\mp}$. Strikingly, they wouldn’t agree on the order of magnitude of the cutoffs. Alice’s correlation length would be large, and she’d assume $\frac{\Lambda}{\Lambda_1}\ll\frac{\Lambda_1}{\Lambda’}$. On the other hand, Bob would assume a small correlation length, so will think that $\frac{\Lambda}{\Lambda_1}\gg\frac{\Lambda_1}{\Lambda’}$. This might not be a contradiction. We assumed that Alice and Bob are widely separated in the RG hierarchy, with
\be
\Lambda’\ll\Lambda_A\ll\Lambda_1\ll\Lambda_B\ll\Lambda~,
\ee
and that each has integrated out the modes that the other resides in. They are both non-deterministic observers, and we assume that only a deterministic observer that hasn’t integrated anything out can compare Alice and Bob’s perspective to each other.

Alternatively, let’s assume that Alice and Bob’s fine tuning is synchronized. Perhaps Bob is speculating about observers like Alice, and is trying to answer the question, ``If the field I am studying is, to some other observer, a UV field with $\mu_0\approx\Lambda’$, what would I see if I used the same fine tuning?”. Bob would assume that the cutoffs of the field he is seeing are consistent with Alice’s cutoffs, $\frac{\Lambda}{\Lambda_1}\ll\frac{\Lambda_1}{\Lambda’}$, and would then integrate out the IR modes with $\xi_{+}=\frac{\Lambda’}{\Lambda_1}$. This would not contradict, for example, Bob’s requirement of a small rescaled correlation length in eq.~(\ref{eq: IR gaussian fixed point trivial bounds condition}). Bob would see a different fixed point than Alice, with
\be
\mu_{-}^{\mathrm{Alice}}\approx\Lambda’_{-}=\frac{\Lambda\Lambda’}{\Lambda_1}~~~\mathrm{and}~~~\mu_{+}^{\mathrm{Bob}}\approx\Lambda’\frac{\Lambda’}{\Lambda_1}\ll\Lambda’_{+}~,
\ee
and a different propagator if we compare eq.~(\ref{eq: Feynmann propagator}) for Alice with eq.~(\ref{eq: IR propagator}) for Bob.

%---------------------------------------------------------------------------------------------------------------------------------------

\paragraph{Active Picture.} Let’s now discuss the active picture, in which Alice studies one field with RG and an entirely different field with IRG. Her RG hierarchy might look like this:
\be
\Lambda’\ll\Lambda_1<\Lambda_2\ll\Lambda~.
\label{eq: active hierarchy}
\ee
For the UV field, we integrate out modes between $(\Lambda_2,\Lambda)$, while for the IR field, we integrate out modes between $(\Lambda’,\Lambda_1)$. Alice will then be working with \emph{two} Planck’s constants, which, in the notation of section \ref{subsec: classical limit}, would be called $\hbar_{2’}$ for the UV field and $\hbar_{1}$ for the IR field, with corresponding expressions for the renormalized $\gamma$ and renormalized speed of light. We assume that the renormalization would be different for the UV and IR fields, but we don’t discuss this further. Each field will have different causal structure because the propagators of UV and IR fields are different.
Finally, say that Alice is studying the UV field at a Gaussian fixed point. We can see that she can then consistently study the IR field at either the Gaussian fixed point or at the fixed point at infinity. For Alice’s Gaussian fixed point UV field, she’ll have $\mu_{2'} t_2\ll1$, with $t_2^{-1}\approx\Lambda_2$. This will lead to 
\be
\frac{\Lambda}{\Lambda_2}\ll\frac{\Lambda_2}{\Lambda’}~.
\label{eq: UV active bounds}
\ee
If she chooses the fixed point at infinity for the IR field, this will lead to the condition $\frac{\Lambda_1}{\Lambda’}\ll\frac{\Lambda}{\Lambda_1}$. Taken with eq.~(\ref{eq: UV active bounds}), we get the requirement $\Lambda_1\ll\Lambda_2$. Interestingly, no such requirement exists if Alice were studying the Gaussian fixed point for the IR field, because then the small IR field correlation length only trivially leads to eq.~(\ref{eq: IR gaussian fixed point trivial bounds condition}). Either way, we’ve not yet specified how $\frac{\Lambda_2}{\Lambda_1}$ might relate to, say, $\frac{\Lambda_1}{\Lambda’}$. We speculate that we should be studying theories in which the latter is much larger, as Alice, in studying the IR field, has access to the modes between $(\Lambda_1,\Lambda_2)$ but not modes between $(\Lambda’,\Lambda_1)$.

%--------------------------------------------------------------------------------------------------------------------------------------------------------------

\subsection{What if Alice (or Bob) Uses Both RG and IRG?}
\label{subsection: Alice's renormalization group}

We now place a single observer (who we’ll call Alice) in the RG hierarchy
\be
\Lambda'\ll\Lambda_1\ll\Lambda_2\ll\Lambda~.
\label{eq: RG/IRG hierarchy}
\ee
RG is used to integrate out modes between $(\Lambda_2,\Lambda)$ and IRG is used to integrate out modes between $(\Lambda’,\Lambda_1)$. If RG is applied first, we write this as $IRG_{1’}\cdot RG_{2}(-\beta H)$, where the subscripts indicate the modes that \emph{were} integrated out. Let’s contrast this with the notation below eq.~(\ref{eq: new notation}). On the one hand, we have quantities associated with multiple applications of the renormalization group: fields, Lagrange parameters, including $\hbar$ and $c$, and rescaled cutoffs. For these quantities, subscripts indicate which modes haven’t yet been integrated out. So in the non-interacting theory, all modes separate like in eq.~(\ref{eq:mode seperation}) and we will be left with an integral over $\phi_{21}$ for the hierarchy in eq.~(\ref{eq: RG/IRG hierarchy}). On the other hand, quantities associated with only a single application of the renormalization group – renormalization group elements and scaling factors – will have their subscripts represent modes that were integrated out. So the scaling factor $\xi_{1’}$ will be associated with application of $IRG_{1’}$ where the low energy modes below $\Lambda_1$ were integrated out.

We want to know what kind of QFT will emerge when RG and IRG are applied to the hierarchy in eq.~(\ref{eq: RG/IRG hierarchy}). We also want to know if successive applications commute,
\be
IRG_{1’}\cdot RG_{2}(-\beta H)=RG_{2}\cdot IRG_{1’}(-\beta H)~.
\label{eq: commuting RG and IRG}
\ee
That is, do we get the same QFT regardless of what we apply first in the RG hierarchy? Obviously, both will end up with a partition function with a functional integral over $\phi_{21}$. But in section \ref{subsec: classical limit}, we saw a qualitatively different QFT for Alice, who studied $RG_{1A}\cdot RG_{1} (-\beta H)$, than for Wigner, who studied $RG_{A}$. The former had an effective Planck’s constant (and speed of light) while the later did not. So we’d like to directly check if the two sides of eq.~(\ref{eq: commuting RG and IRG}) can stand for the same QFT. To help us, we will stick to the following criteria to generally determine if two distinct applications of the renormalization group have led to the same emergent QFT:\\
\\{\em Say that we act on a Hamiltonian with two RG operations, $\mathcal{O}_1$ and $\mathcal{O}_2$, subject to the postulates of section \ref{subsec: QFT_postulates}. Then the two
renormalized Hamiltonians $\mathcal{O}_1 H$ and $\mathcal{O}_2 H$ lead to the same effective QFT if they: a) have the same
fixed point; b) have the same renormalized speed of light; c) have the same Planck’s constant; and, d) can consistently take measurements at the same time t.}\\

%---------------------------------------------------------------------------------------------------------------------------------------

\paragraph{First one way.} We begin with $IRG_{1’}\cdot RG_{2}(-\beta H)$:
\bea
IRG_{1’}\cdot RG_2 (-\beta H) &=&IRG_{1’}\left[\frac{1}{2\hbar_{2’}}\int^{\Lambda_2\xi_2}_{\Lambda’\xi_2}\frac{\mathrm{d}^4k}{(2\pi)^4}\phi_{2'}(k)\phi_{2'}(-k)
[k_0^2+c_0^2\vec{k}^2+\mu_{2’}^2c_{2’}^4]\right]\notag \\
&=&\frac{1}{2\hbar_{21}}\int^{\Lambda_2\xi_2\xi_{1’}}_{\Lambda_1\xi_{1’}}\frac{\mathrm{d}^4k}{(2\pi)^4}\phi_{21}(k)\phi_{21}(-k)
\left[k_0^2+c_0^2\vec{k}^2+\mu_{21}^2c_{21}^4\right]~.
\label{eq: RG then IRG}
\eea
We’ve yet to specify the scaling terms $\xi_{2}$ and $\xi_{1’}$. We explicitly write out the initial application of $RG_2$, for which $\hbar_{2’}=T\gamma_{2’}\epsilon^{-1}$ and $c=c_0\gamma_{2’}^{\frac{1}{4}}$. We presume that the application of IRG will separately renormalize the field, so we have two gamma’s in $\hbar_{21}=T\gamma_{2’}\gamma_{21}\epsilon^{-1}$  and the corresponding expression for the speed of light. The renormalized masses are
\be
\mu_{2’}=\mu_0\xi_{2}\gamma_{2’}^{-1}~~~\mathrm{and}~~~\mu_{21}=\mu_0\xi_{2}\xi_{1’}\gamma^{-1}_{2’}\gamma^{-1}_{21}~.
\label{eq: renormalized masses for IRG/RG}~.
\ee
To get from the 1st equality of eq.~(\ref{eq: RG then IRG}) to the 2nd, we also needed the self consistency condition
\be
\Lambda’\xi_2\ll\Lambda_1~.
\label{eq: consistency condition, IRG RG}
\ee

For the non-interacting theory, the requirements to have the same Planck’s constant and speed of light mentioned in the criteria below eq.~(\ref{eq: commuting RG and IRG}) are trivially satisfied, so we set all $\hbar$, $c$, and $\gamma$ equal to 1 in what follows.  We’d like to study the Gaussian fixed point with eq.~(\ref{eq: commuting RG and IRG}) and one way to do this is to fine tune with $\mu_0\approx\frac{\Lambda_1\Lambda_2}{\Lambda}$ and to set the scaling factors to $\xi_2=\frac{\Lambda}{\Lambda_2}$ and $\xi_{1’}=\frac{\Lambda’}{\Lambda_1}$. We have
\bea
IRG_{1’}\cdot RG_2 (-\beta H)
&=& IRG_{1’}\left(\frac{1}{2}\int^{\Lambda}_{\frac{\Lambda\Lambda’}{\Lambda_2}}\frac{\mathrm{d}^4k}{(2\pi)^4}\phi_{2'}(k)\phi_{2'}(-k)[k^2+\mu_{2’}^2]\right) \notag \\
&=& \frac{1}{2}\int^{\frac{\Lambda\Lambda’}{\Lambda_1}}_{\Lambda’}\phi_{21}(k)\phi_{21}(-k)[k^2+\mu_{21}^2]~.
\eea
What’s the fixed point for $RG_2 (-\beta H)$? From the consistency condition eq.~(\ref{eq: consistency condition, IRG RG}) and our choice of scaling, we have a condition on the bounds,
\be
\frac{\Lambda’}{\Lambda_1}\ll\frac{\Lambda_2}{\Lambda}
\label{eq: bounds condition, IRG RG}~,
\ee
and then, with eq.~(\ref{eq: renormalized masses for IRG/RG}), we have $\mu_{2’}\approx\Lambda_1$. Note that this means that
\be
\Lambda’_{2’}\ll\mu_{2’}\ll\Lambda_{2’}~,
\ee
which is a fixed point we haven’t encountered yet.

As advertised, the fixed point for $IRG_{1’}\cdot RG_2 (-\beta H)$ is Gaussian: $\mu_{21}=\Lambda’=\Lambda’_{21}$~. We’d also like to have $\mu_0 t\ll1$ and $\mu_{21}t\ll1$ as with the Gaussian fixed point in section \ref{subsection: hbar, c, and postulate 3}. The first condition requires $t^{-1}\gg\frac{\Lambda_1\Lambda_2}{\Lambda_2}$, and for consistency with the next section, we choose
\be
t^{-1}\approx\Lambda_2~.
\label{eq: time for IRG RG}
\ee
The second condition on the time is the trivially satisfied, with $\frac{\Lambda’}{\Lambda_2}\ll1$.\\

%---------------------------------------------------------------------------------------------------------------------------------------

\paragraph{Then the other way.} We now integrate out the IR modes before the UV ones:
\bea
RG_{2}\cdot IRG_{1’} (-\beta H)
&=& RG_2\left(\frac{1}{2}\int_{\Lambda_1\xi_{1’}}^{\Lambda\xi_{1’}} \frac{\mathrm{d}^4k}{(2\pi)^4}\phi_1(k)\phi_1(-k)[k^2+\mu_1^2]\right)\notag \\
&=& \frac{1}{2}\int^{\Lambda_2\xi_{2}}_{\Lambda_1\xi_{1’}\xi_2}\frac{\mathrm{d}^4k}{(2\pi)^4} \phi_{21}(k)\phi_{21}(-k)[k^2+\mu_{21}^2]~.
\eea
To get to the 2nd equality, we needed, for consistency,
\be
\Lambda\xi_{1’}\gg\Lambda_2~.
\label{eq: consistency condition, RG IRG}
\ee
The renormalized masses are now
\be
\mu_1=\mu_0\xi_{1’}~~~\mathrm{and}~~~\mu_{21}=\mu_1\xi_2~.
\ee
To get the Gaussian fixed point with $RG_{2}\cdot IRG_{1’} (-\beta H)$, we fine tune to $\mu_0\approx\Lambda_1$. We also set the scaling to $\xi_{1’}=\frac{\Lambda’}{\Lambda_1}$ and $\xi_2=\frac{\Lambda}{\Lambda_2}$. With this, we also hit a Gaussian fixed point with $IRG_{1’}(-\beta H)$ with $\mu_{1’}\approx\Lambda’$.

The condition in eq.~(\ref{eq: consistency condition, RG IRG}) becomes
\be
\frac{\Lambda’}{\Lambda_1}\gg\frac{\Lambda_2}{\Lambda}
\label{eq: bounds condition, RG IRG}~,
\ee
which is different than eq.~(\ref{eq: bounds condition, IRG RG}). We presume that we can still have the same QFT if the bounds and fine tunings are different, so long as the criteria below eq.~(\ref{eq: commuting RG and IRG}) are fulfilled. We saw a similar disagreement in bound limits when discussing the passive picture, below eq.~(\ref{eq: fine tuning 1, passive picture}). Looking at the condition on the correlation length for a Gaussian fixed point, we have $t\Lambda_1\ll1$ so can again choose a time like in eq.~(\ref{eq: time for IRG RG}). The condition on the rescaled correlation length, $t \mu_{21}\ll1$, now leads to a further condition on the bounds, $\frac{\Lambda}{\Lambda_2}\ll\frac{\Lambda_2}{\Lambda’}$, which is not in conflict with eq.~(\ref{eq: bounds condition, RG IRG}). With the fixed point and time the same (and $\hbar$ and $c$ set to 1), we have established eq.~(\ref{eq: commuting RG and IRG}).\\

%---------------------------------------------------------------------------------------------------------------------------------------

\paragraph{The Renormalization Group as a Group.} Because we deal with the non-interacting field, how $\hbar$ is renormalized when both UV and IR modes are integrated out is beyond the scope of this paper. If the heuristic of section \ref{subsec: QFT_postulates} was applicable, $\hbar$ would decrease as both UV and IR modes are integrated out. We’ve assumed that this is because an observer studying such a field lacks information on the integrated-out modes. But it should ultimately be the field theory that tells us how $\hbar$ is renormalized, so let’s also imagine an alternative scenario where $\hbar$ decreases with RG but \emph{increases} with IRG. It would be harder to reach the classical limit with such a field, as more UV modes would have to be integrated out than would otherwise to get a small enough $\hbar_{\mathrm{eff}}$.

We can also imagine ``limit cycles” with fixed points from multiple applications of RG and IRG. Take, for example, the hierarchy
\be
\Lambda’\ll\Lambda_1\ll\Lambda_2\ll\Lambda_3\ll\Lambda
\ee
in the non-interacting scalar theory and say that we integrate out modes in the order $RG_{32}\cdot IRG_{1’}\cdot RG_3$. Say that $RG_3$ gives a Gaussian fixed point and then $IRG_{1’}\cdot RG_3$ takes us to some other fixed point, say the one below the lower rescaled cutoff like in eq.~(\ref{eq:mu_+ for gaussian fixed point}). Perhaps with the final application of $RG_{32}$ we are brought back to a Gaussian fixed point. In addition, Planck’s constant decreases as modes are integrated out with $\Lambda_3$ and $\Lambda_2$ and increases (with our new heuristic) as $\Lambda_1$ does. We could then imagine that to leading order, $\hbar_{3’}$ flows like $\hbar_{21}$. Taken together and with the criteria below eq.~(\ref{eq: commuting RG and IRG}) in mind, we might conclude that we get the same QFT with $RG_{32}\cdot IRG_{1’}\cdot RG_3 (-\beta H)$ as we would with just $RG_3 (-\beta H)$. The speculation is that the renormalization group is now acting like a group, even though we are certainly not getting back any modes we have integrated out: the effective Hamiltonian for $RG_{32}\cdot IRG_{1’}\cdot RG_3 (-\beta H)$ will involve only the unintegrated modes $\phi_{21}$. We do not have an example in mind that would realize these possibilities and only bring them up here as a thought experiment; an example of what might be possible if IRG is used with RG.

%----------------------------------------------------------------------------------------------------------------------------------------------

\section{Summary and Conclusion}

We’ve argued that quantum mechanics, and special relativity, can be viewed as emergent phenomena. The price is an expanded RG ontology, in which one observer’s UV field is a different observer’s IR field, and where QFT takes on different guises depending on where an observer is located in the RG hierarchy. The theory is realist, so all perspectives are subsumed under that of a deterministic observer who has complete knowledge about modes of the field they are studying, and doesn’t have to integrate anything out, or, if they do, what they integrate out is paltry in comparison to someone who sees a field as a QFT. We’ve tried to develop a process for creating QFT perspectives linked to the procedures of the renormalization group, with the aim of freeing us to study QFTs up and down an RG hierarchy. We hope that the idea that one person’s high energy physics is another person’s observable universe, where the latter perspective is located at a higher energy then the former imagines exists, ``democratizes” physics at vastly different scales. We are used to doing RG by imagining that the high energy modes are very complicated and that, with application of RG flow, we get simple physics at low energy. Why not imagine a perspective where the opposite is true? And why not imagine that both perspectives describe the same object?

This paper leaves a lot unexplored, so we briefly discuss potential avenues of research here. Just with the non-interacting field, it might be helpful to study this formalism with the grand canonical ensemble. We’ve also not discussed how quantum statistical mechanics is emergent in RG ontology. If $\hbar$ absorbs the temperature at the fixed point, how do you get back both $\hbar$ and T? We presume that this is related to the correspondence between Euclidean QFT and quantum statistical mechanics, and that we should still start with a vacuum theory to define $\hbar\propto T_c$, but we do not explore this connection here. We should also calculate $A$ from eq.~(\ref{eq: functional measure}), which we assumed proportional to Planck’s constant because of the connection between many particle quantum mechanics and field theory.

We’ve mentioned the need for interactions several times in this paper. Can we reproduce behavior of the renormalized Lagrange coefficients and beta functions with an $\hbar$ and $c$ that obey our renormalization heuristic? What would renormalized coefficients and flow of $\hbar$ be for an IR field with interactions? We’ve discussed the cutoff rescaling criteria and how it leads to fixed points other than the Gaussian for the non-interacting theory, like the so called ``massless fixed point”. Can this be generalized with interactions? 

Our discussions of RG invariants, in subsection \ref{subsec: RG invariants}, and on RG as a group, at the end of section \ref{subsection: Alice's renormalization group}, were purely speculative. In general, we’d like to put the RG hierarchy and things like the cutoff rescaling criteria on firmer mathematical footing. There is also much that we left out of our discussion on quantum foundations, from thought experiments like Wigner’s enemy to a deeper discussion of entanglement and decoherence. The hope for this paper is that it serves as a jumping off point for further research.

It is also not clear whether something like the observable universe can really be viewed as some sort of IR field imbedded in an RG hierarchy with the low energy modes integrated out. Such an object would certainly not be a non-interacting scalar field! This would ostensibly place modes larger than the observable universe as far away from us as the modes presumed to affect high energy fields, which this formalism assumes are at scales smaller than the Planck length. And can we apply the formalism of RG ontology to gravity? Perhaps the process of starting with a non-relativistic classical field in the RG hierarchy will lead to a renormalized metric. We certainly cannot answer any of these questions in this paper.

It is a cliché that there is a zoo of quantum mechanical interpretations. The abundance can deter progress when, say, advances in QFT aren’t incorporated into interpretations that rely on explanations of non-relativistic QM\cite{wallace2022skybluereasonsquantum}. We add this interpretation to the pile in a more hopeful idiom, in the belief that each new interpretation adds a little to what we know and gets us closer to a world that has been, somehow, obscured from view.

\section*{Acknowledgements}

Many thanks to Andrea Erdas, for your notes on the manuscript and for the illuminating discussions we've had as this work has progressed. Thanks also to Adrian Dumitru for discussing the manuscript with me and for your support.

\bibliography{References_2025}

\typeout{get arXiv to do 4 passes: Label(s) may have changed. Rerun}

\end{document}